# Nanoscale control of rewriteable doping patterns in pristine graphene/boron nitride heterostructures

Jairo Velasco Jr.[1,5,†], Long Ju[1,†], Dillon Wong[1,†], Salman Kahn[1], Juwon Lee[1], Hsin-Zon Tsai[1], Chad Germany[1], Sebastian Wickenburg[1], Jiong Lu[1,‡,Δ], Takashi Taniguchi[4], Kenji Watanabe[4], Alex Zettl[1,2,3], Feng Wang[1,2,3] and Michael F. Crommie[1,2,3],*

[1]*Department of Physics, University of California, Berkeley, California 94720, USA*

[2]*Materials Sciences Division, Lawrence Berkeley National Laboratory, Berkeley, California 94720, USA*

[3]*Kavli Energy NanoSciences Institute at the University of California, Berkeley and the Lawrence Berkeley National Laboratory, Berkeley, California 94720, USA*

[4]*National Institute for Materials Science, 1-1 Namiki, Tsukuba, 305-0044, Japan*

[5] *Department of Physics, University of California, Santa Cruz, California 95064, USA*

[†] *These authors contribute equally to this manuscript.*

*\* Email:crommie@berkeley.edu*

[‡] *Present address: Department of Chemistry, National University of Singapore, 3 Science Drive 3, Singapore 117543,Singapore*

[Δ] *Present address: Centre for Advanced 2D Materials and Graphene Research Centre, National University of Singapore, 6 Science Drive 2, Singapore 117546*

## Abstract:

Nanoscale control of charge doping in two-dimensional (2D) materials permits the realization of electronic analogs of optical phenomena, relativistic physics at low energies, and technologically promising nanoelectronics. Electrostatic gating and chemical doping are the two most common methods to achieve local control of such doping. However, these approaches suffer from complicated fabrication processes that introduce contamination, change material properties irreversibly, and lack flexible pattern control. Here we demonstrate a clean, simple, and reversible technique that permits writing, reading, and erasing of doping patterns for 2D materials at the nanometer scale. We accomplish this by employing a graphene/boron nitride (BN) heterostructure that is equipped with a bottom gate electrode. By using electron transport and scanning tunneling microscopy (STM), we demonstrate that spatial control of charge doping can be realized with the application of either light or STM tip voltage excitations in conjunction with a gate electric field. Our straightforward and novel technique provides a new path towards on-demand graphene pn junctions and ultra-thin memory devices.

Local control of charge doping in 2D systems allows for the study of new classes of phenomena that include electronic lensing[1,2], Klein tunneling[3,4], ultra-thin light emitting diodes,[5-7] and *in-situ* tunable plasmonic platforms[8]. The fabrication of nanostructures exhibiting such doping, however, requires sophisticated lithography procedures that compromise sample quality and do not yield flexible control of doping patterns[9-14], thus hampering advancement in these areas. Some progress has been made in patterning buried GaAs/AlGaAs heterojunctions via a lithography-free conducting AFM tip-based technique[15], although only for a single dopant polarity. For more open graphene/BN heterostructures it has been shown that illumination via visible light causes defect charge migration within the BN layer, thus shifting the graphene charge neutrality point (CNP)[16] and inducing doping. Currently, however, there is no understanding of how such defect charge migration varies spatially at the nanoscale and whether it might be exploited as a tool for developing new 2D devices.

Here we address these issues directly via a new scheme that combines optical and local electric field excitation of defect charge in a BN insulator overlaid with graphene. This permits us to write, read, and even erase doping patterns at the nanoscale. Unlike other tip-based patterning methods[15], this new technique allows direct wavefunction mapping of 2D devices and control over local doping polarity. Fig. 1 shows two different schemes that we used for creating our rewritable doping patterns in a graphene/BN heterostructure supported by an SiO$_2$/Si substrate and contacted by gold electrodes. A backgate voltage $V_g$ applied to the doped Si substrate permits global tuning of the graphene doping level, while local modification is achieved by excitation (via photons or DC electric field) of electrons (holes) from donor-like (acceptor-like) defects in BN. The liberated charge leaves behind ions that locally gate the graphene layer above (this is equivalent to doping for graphene). Figs 1a and 1b depict this process for a donor-like defect via optical excitation (Fig. 1a) and tip-induced electric field (Fig. 1b).

We first explore microscopic spatial fluctuations in defect charge doping induced by light. Fig. 2a shows the result of transport measurements (σ vs. $V_g$) taken on our graphene/BN heterostructure after different light exposures. The response of the device prior to light exposure is shown in the red trace and is referred to as State 1. Here the dip in σ (which marks the CNP) is located at $V_g \sim 0$ V. The narrow width of the CNP dip is evidence that the sample starts off with a low amount of spatial fluctuations in the underlying charge landscape of the BN substrate[17]. We then set the gate to $V_g = -38$ V and exposed the sample to light for ~45 seconds ($\lambda = 405$ nm, power density ~10 µW µm$^{-2}$ ). The σ($V_g$) curve measured immediately afterwards (with the light off) is shown in the green trace and is referred to as State 2. Here we see that the CNP dip is much wider and has shifted to $V_g \sim -15$ V. Next we exposed the device to light for a significantly longer period (~10 minutes) with the gate again held at $V_g = -38$ V. The σ($V_g$) measurement performed afterwards is shown as the blue trace and is referred to as State 3. The CNP dip is now seen to have narrowed to its original width, but it is shifted to $V_g = -38$ V, precisely the value of the gate voltage during the light exposure.

The differences in shape between the three σ($V_g$) curves provide evidence of a microscopic doping pattern induced by light exposure. The broadened width for State 2 indicates that it has a different doping landscape compared to State 1 — an enhancement in charge inhomogeneity[18]. We attribute this change to light-induced redistribution of charged impurities in the BN substrate. Optically excited electrons from defect sites (that are distributed randomly within the BN crystal) migrate both to the graphene and to other randomly distributed defects under the influence of the gate electric field and homogenous illumination. Short light exposure (i.e., where defect charge migration is insufficient to fully screen the gate potential) leads to a disordered distribution of the overall charged defects that increase charge inhomogeneity within the graphene, thus widening $\sigma(V_g)$ (as seen in State 2). Upon further light exposure, the charged defects form a more even distribution in order to fully screen the gate-induced electric field in the BN, thus "erasing" the charge inhomogeneity induced by the shorter light exposure (as evidenced by State 3). Closer inspection of our transport measurements for States 1-3 further support the emergence of a microscopic doping pattern induced by light. This can be seen in Fig. 2b where the electron branch of the $\sigma(V_g)$ curves are replotted as $\sigma(n)$ with $n$ on a logarithmic scale. Here the average amplitude of the charge inhomogeneity in graphene is measured by the point of intersection of the two lines extending from the constant sloped regions[19]. This is marked by colored arrows for each curve. For State 2 the charge inhomogeneity amplitude is larger by an order of magnitude than States 1 and 3, and so reveals a doping landscape patterned by light.

In order to directly visualize the nanoscale characteristics of this doping pattern, we performed STM measurements on our graphene/BN heterostructure after different optical treatments. Figs. 2c-e show differential conductance (d$I$/d$V$) maps obtained at constant sample bias $V_s$ and gate voltage $\tilde{V}_g$ ($\tilde{V}_g$ denotes the gate voltage applied during a d$I$/d$V$ measurement while $V_g$ denotes the gate voltage applied during exposure of the sample to light or a voltage pulse). These measurements were performed at the same location both before and after light exposures similar to those discussed in Figs 2a-b. All maps exhibited a 7 nm moiré pattern, indicating a clean graphene/BN interface. Before any light exposure (Fig. 2c) we observe low spatial variation in the d$I$/d$V$ intensity, except for a single charged defect visible as a red dot in the lower right quadrant of the map[20]. Next we retracted the tip, set $V_g$ = -15 V, and exposed the device to a short light exposure of 20 seconds (similar to how State 2 was prepared in Fig. 2a). As seen in Fig. 2d, the resulting d$I$/d$V$ map exhibits significant new charge inhomogeneity, observable as an irregularly shaped red area in the top right portion of the map. We then retracted the tip, set $V_g$ = -15 V, and performed a long light exposure for 15 minutes on the device (similar to how State 3 was prepared in Fig. 2a). As seen in Fig. 2e, this erases the charge inhomogeneity and reduces it to the same level as observed for the pristine sample in Fig. 2c. The sequence of observations in Figs 2(c-e) was replicated at numerous locations with many tips and with different devices.

The changes observed in our d$I$/d$V$ maps provide a direct visualization of how the doping landscape can be tuned by light excitation. Because variations in d$I$/d$V$ are proportional to changes in the electronic local density of states, d$I$/d$V$ maps taken at a fixed bias near graphene's Dirac point (DP) reflect spatial variation in the DP energy[21, 22]. Such images can be converted to a charge density fluctuation amplitude δ$n$ in a straightforward way (see supporting figure S1). The small variations in d$I$/d$V$ intensity of Fig. 2c (the image prior to light exposure) indicate a small δ$n$ (~$10^9$ cm$^{-2}$), consistent with previous STM studies of graphene/BN heterostructures[23, 24]. The charge density fluctuations seen in Fig. 2d (after a short light exposure), however, reflect much larger charge fluctuations (δ$n$ ~ $10^{10}$ cm$^{-2}$ across the image). This is consistent with the broadened CNP feature observed in the transport measurement of State 2 (Fig. 2a). The reduction of charge inhomogeneity seen in the spatial map of Fig. 2e after long light exposure is consistent with the narrowing of the CNP feature observed via a transport measurement of State 3 after similar processing (Fig. 2a). Both the d$I$/d$V$ maps and the transport data support a light-induced charge doping mechanism where optical excitation frees defect charge in the insulating layer that then migrates in response to the gate electric field. Here nanoscale inhomogeneity arises from spatial disorder in the BN defect density, which causes transient fluctuations in the local graphene charge carrier density as the defect charge rearranges itself to screen the backgate field.

These light-induced doping patterns emerge at arbitrary positions because the BN defects (the source of the migrating charge within the BN) are randomly located throughout the BN crystal. A natural question to ask is whether this defect-mediated charge migration process might be harnessed and controlled locally with more precision. A recent experiment found that individual defects in BN can be ionized by STM tip voltage pulses[20], but that experiment was performed in the absence of a gate electric field (unlike the experiments discussed here) and detected no net charge exchange between graphene and the insulating substrate. To test the effect of a local ionizing potential on defect charge migration in the presence of a gate electric field, we applied tip voltage pulses to a graphene/BN device using the experimental setup sketched in Fig. 1b. Fig. 3a shows d$I$/d$V$ spectra obtained with $\tilde{V}_g = 0$ V after applying $V_s = 5$ V tip voltage pulses while holding $V_g$ at different values (the d$I$/d$V$ spectra were obtained at the same location the tip pulses were applied). The red trace shows the reference spectrum measured before application of any tip pulses. Here we observe a ~130 mV gap-like feature at the Fermi energy[24, 25] that is known to arise due to phonon-mediated inelastic tunneling[25]. To the right of this inelastic tunneling feature is a dip (black arrow) that marks the DP. Because the DP lies to the right of the Fermi energy ($V_s = 0$ V), we see that this region of the surface has residual p-doping (~5 x$10^{11}$ cm$^{-2}$) at zero gate voltage. The yellow, green, and blue traces show the d$I$/d$V$ spectra measured after applying tip pulses lasting 30 sec. with the gate voltage set respectively to $V_g = -10$ V, -20 V, and -30 V (the height of the tip in each case was approximately 1.5 nm away from the surface – see supplementary information for details). As shown by the black arrows, the DP shifts down in energy as each tip pulse is delivered with a more negative gate voltage. The sample is seen to locally change from p-type doping to n-type doping after the first pulse, and then to become more heavily n-doped after each pulse. This behavior is consistent with the local BN charge landscape becoming increasingly positively charged after tip pulses performed at increasingly negative gate voltages. Reversing the polarity of the gate field, while leaving everything else the same, results in local graphene doping with the exact opposite polarity (see supporting figure S2).

We were able to gain insight into the spatially varying dopant landscape that results from a tip pulse by performing d$I$/d$V$ imaging of the area beneath the STM tip both before and after a tip pulse. All maps exhibited a 7 nm moiré pattern, indicating a clean graphene/BN interface. Fig. 3b shows a d$I$/d$V$ map of a patch of graphene right before performing a tip pulse. It contains a number of point-like defects due to charge centers in the BN layer[20], but otherwise exhibits a smooth charge landscape. We next brought the STM tip to the top right corner of this region and applied a tip voltage pulse while holding the gate voltage at $V_g$ = -20 V. Fig. 3c shows a d$I$/d$V$ map of the same region after applying the tip pulse. The most striking feature in the d$I$/d$V$ map after the tip pulse is the emergence of a red disk region in the upper right quadrant of the map, which also exhibits a darkened halo around the perimeter. Although only one quadrant is shown, the new red region exhibits rough circular symmetry. The altered charge landscape is stable at T = 5 K long after the pulse has been applied, but it can be erased by application of an identical tip pulse with the gate voltage held at $V_g$ = 0 V. Fig. 3d shows a d$I$/d$V$ image of the same graphene patch after application of such an "eraser" pulse. The altered red disk is now completely gone and the graphene is returned to its pristine state.

The new charge doping landscape induced by the tip pulse can be explained by a combination of field-induced defect ionization and charge diffusion within the BN insulator (see supporting information for more discussion on BN defects). The strong electric field of the tip pulse penetrates through the gated graphene into the insulator region (previous studies have also shown similar electric field penetration through gated graphene[26]), causing a strong potential gradient around BN defects and resulting in enhanced defect field emission. When the gate is on during a tip pulse, the gate electric field causes released electrons to drift either into the graphene electrode ($V_g < 0$, resulting in a positive space charge layer in the BN) or away from the graphene electrode ($V_g > 0$, resulting in a negative space charge layer in the BN). Pulses applied with $V_g = 0$ V allow charge to freely diffuse and recover the initial state of graphene. The net result is that both p-type and n-type doping profiles can be written and erased in pristine graphene/BN with a spatial resolution determined by the potential gradient surrounding an STM tip. For example, the red region in Fig. 3c is n-doped graphene while the blue region surrounding it is p-doped graphene (see supporting figure S3), and the boundary between these two regions defines a re-writable nanoscale p-n junction (we have also performed transport measurements that clearly demonstrate the creation of a graphene p-n junction using our tip-doping technique (Fig. S4)).

The strong similarities in phenomenology between both the light-induced and tip-pulsing-induced doping modalities described here suggest that they might even be combined. To demonstrate this compatibility, we show how a localized n-doped region can be "written" into a p-doped background and then erased by optical excitation. Fig. 4a shows a d$I$/d$V$ map of an n-doped region (~200 nm across) created through application of a tip pulse at the center of the image, while holding the gate voltage at $V_g = -20$ V. Light-induced erasure of this pattern was subsequently performed by exposing the sample to light for 30 min., while holding the gate voltage at $V_g = 0$ V. Fig. 4b shows a d$I$/d$V$ map of the same region after the erasure process. The red n-doped region is clearly absent, demonstrating that light exposure can erase doping patterns fabricated via tip pulses. These observations show that combined tip and optical-induced charge doping is a flexible scheme for creating rewritable p-n junctions of arbitrary geometry.

In the present work we have demonstrated writing, reading, and erasing of artificial doping patterns at the nanoscale in pristine graphene/BN heterostructures. Our local patterning method uses light or voltage pulses to liberate charge from BN defects, which is then guided by the gate electric field. This simple and clean defect-based lithography method provides flexibility both for writing and erasing operations. The ultimate spatial resolution of this technique will depend on the geometry of the probe tip, which could be enhanced through the use of high aspect-ratio structures such as nanotubes. Our charge-doping control scheme may also be employed on other types of low-dimensional heterostructures, such as transition metal dichalcogenides and flexible memory cells[27, 28].

## Methods:

Our samples were fabricated using a transfer technique developed by Zomer and colleagues[29] that employs standard electron-beam lithography. We used high purity BN crystals synthesized by Taniguchi *et al.*[30] and exfoliated to 60-100 nm thickness, with $SiO_2$ thicknesses of 300 or 285 nm used as the dielectric for electrostatic gating. Monolayer graphene was exfoliated from graphite and deposited onto methyl methacrylate (MMA) polymer and transferred onto BN (that was also annealed before graphene was transferred onto it) supported by an $SiO_2$/Si wafer. Completed devices were subsequently annealed in flowing $Ar/H_2$ forming gas at 350°C, and their electrical conductance was measured with a standard ac voltage bias lock-in technique with a 50 µV signal at 97.13 Hz. A fiber-based super continuum laser was guided into a ($T = 77$ K) cryostat through an optical window for the transport measurements presented in Fig. 2a. Samples that exhibited bipolar transport within a gate voltage range of -30 V to 30 V were transferred into our Omicron ultra-high vacuum (UHV) low temperature STM. A second anneal was then performed for several hours at ~300°C and $10^{-11}$ torr before moving the device into the STM chamber for measurements at $T = 5$ K. Before all STM measurements our platinum iridium STM tip was calibrated by measuring the Shockley surface state of an independently cleaned Au(111) crystal. All STM images were acquired in a constant current mode with $-V_s$ applied to the STM tip relative to a grounded sample. All scanning tunneling spectroscopy (STS) measurements were obtained by lock-in detection of the a.c. tunnel current induced by a modulated voltage (6-10 mV at 613.7 Hz) added to $V_s$. A diode laser (405 nm wavelength) and an Ar ion laser (454-514 nm wavelength) that are external to the STM were used for light exposure before the acquisition of the $dI/dV$ maps presented in Figs 2c-e.

**Fig. 1**: **Scheme for creating rewritable nanoscale doping patterns on graphene/BN heterostructures.** (a) Light-based scheme: graphene/BN heterostructure supported by $SiO_2$/Si substrate. The graphene is contacted by gold electrodes and a backgate voltage $V_g$ is applied to the doped Si substrate. A doping pattern is induced in the sample by exposing it to light for a short period while holding $V_g \neq 0$ V. (b) Tip-based scheme: device structure is the same as in (a). Here a doping pattern is induced by applying a voltage pulse to the tip while holding $V_g \neq 0$ V. Erasing is accomplished by either light-based or tip-based excitation while holding $V_g = 0$ V.

**Fig. 2**: **Nanoscale charge inhomogeneity controlled with light.** (a) $\sigma(V_g)$ curves obtained after different light exposures. Red trace: pristine sample. Green trace: acquired after a short light exposure (45 seconds) while holding $V_g = -38$ V. Blue trace: acquired after extended light exposure (10 minutes) while holding $V_g = -38$ V. (b) $\sigma(n)$ curves (with $n$ plotted on a logarithmic scale) indicate the amplitude of charge inhomogeneity, $\delta n$ (denoted by the colored arrows) from (a). Data from (a-b) were acquired from a device with a 9 μm width and 2 μm source-drain separation. (c) d$I$/d$V$ map of pristine graphene/BN ($I$ = 0.2 nA, $V_s$ = -0.25 V, $\tilde{V}_g$ = 5 V). $\tilde{V}_g$ denotes the gate voltage applied during the acquisition of a d$I$/d$V$ map. (d) d$I$/d$V$ map of the same region after a short light exposure (20 seconds), while holding $V_g$ = -15 V (acquired with $\tilde{V}_g$ = -10 V and same $I$ and $V_s$ as (c)). (e) d$I$/d$V$ map of the same region after extended light exposure (15 minutes), while holding $V_g$ = -15 V (acquired with $\tilde{V}_g$ = -13 V and same $I$ and $V_s$ as (c)). ($\tilde{V}_g$ is slightly different for each d$I$/d$V$ map so that they could all be acquired at the same global charge density). Data from (c-e) were acquired from a device with a 44 μm width and 39 μm source-drain separation.

**Fig. 3**: **Nanoscale doping patterns controlled with an STM tip voltage pulse.** (a) d$I$/d$V$ spectroscopy of a pristine surface before a tip pulse (red) and after a tip pulse ($V_s$ = 5 V, 30 sec.) for $V_g$ = -10 V (yellow), -20 V (green), and -30 V (blue). Initial tunneling parameters: I = 0.4 nA, $V_s$ = -0.5 V, $\tilde{V}_g$ = 0. The curves are vertically offset for clarity. (b) d$I$/d$V$ map of pristine graphene/BN ($I$ = 0.4 nA, $V_s$ = -0.25 V, $\tilde{V}_g$ = 5 V). (c) d$I$/d$V$ map of the same region after a tip pulse was applied in the corner of the map (location denoted by cross hair), while holding $V_g$ = -20 V ($I$ = 0.4 nA, $V_s$ = -0.25 V, $\tilde{V}_g$ = -10 V). (d) d$I$/d$V$ map of the same region after another tip pulse was applied in the same location, while holding $V_g$ = 0 V ($I$ = 0.4 nA, $V_s$ = -0.25 V, $\tilde{V}_g$ = 5 V). All data were acquired from a device with a 44 μm width and 39 μm source-drain separation.

**Fig. 4**: **Doping patterns written by an STM tip and erased by light.** (a) d$I$/d$V$ map of graphene/BN after applying a tip pulse at the center, while holding $V_g$ = -20 V ($I$ = 0.4 nA, $V_s$ = -0.25 V, $\tilde{V}_g$ = -15 V). (b) d$I$/d$V$ map of the same region after extended light exposure (30 minutes) while holding $V_g$ = 0 V (same tunneling parameters as (a)). All data were acquired from a device with a 44 μm width and 39 μm source-drain separation.

**Acknowledgements:**

The authors thank P. Yu, J. Jung and A. Rubio for stimulating discussions and S. Onishi for help with the scanning electron microscope. This research was supported by the Director, Office of Science, Office of Basic Energy Sciences of the U.S. Department of Energy under contract no. DE-AC02-05CH11231 (sp2 program) (STM imaging and spectroscopy), National Science Foundation grant DMR-1206512 (sample fabrication). D.W. was supported by the Department of Defense (DoD) through the National Defense Science & Engineering Graduate Fellowship (NDSEG) Program, 32 CFR 168a. K.W. and T.T. acknowledge support from the MEXT Japan Elemental Strategy Initiative (synthesis of BN crystals) and JSPS Grant-in-Aid for Scientific Research on Innovative Areas no. 25107004 (characterization of BN crystals). T.T. acknowledges support from a JSPS Grant-in-Aid for Scientific Research on Innovative Areas no. 25106006 (development of high pressure BN synthesis instrumentation). J. L. acknowledges the National Research Foundation, CRP award "Novel 2D materials with tailored properties: beyond graphene" (R-144-000-295-281).

**Author Contributions:**

J.V.J., L.J., and D.W. conceived the work and designed the research strategy. J.V.J. and D.W. performed data analysis. J.V.J., S.K., L.J. and A.Z. facilitated sample fabrication. D.W., J.L. and J.V.J. carried out STM/STS measurements. J.V.J., L.J and S.K. carried out electron transport measurements. L. J. constructed the optical setup. K.W. and T.T. synthesized the h-BN samples. D.W., J.V.J. and L.J. formulated theoretical model with advice from M.F.C. M.F.C. supervised the STM/STS experiments. J.V.J., L.J., D.W. and M.F.C. co-wrote the manuscript. J.V.J. and M.F.C. coordinated the collaboration. All authors discussed the results and commented on the paper.

**Fig. 1**

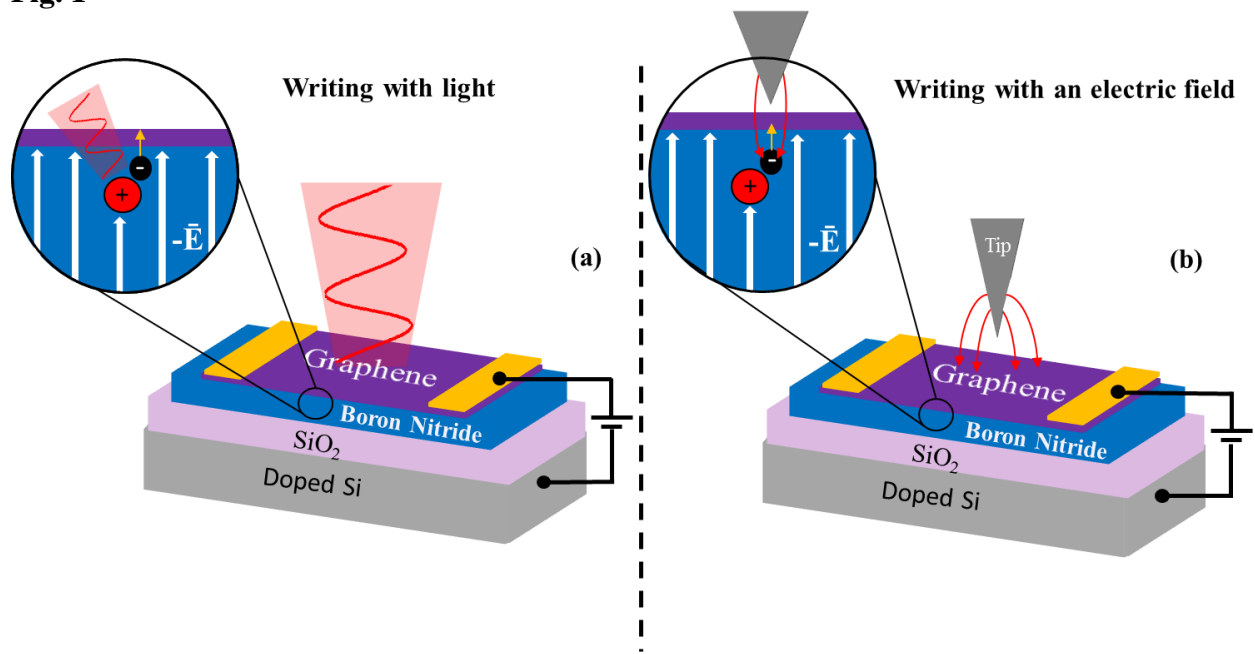

Fig. 2

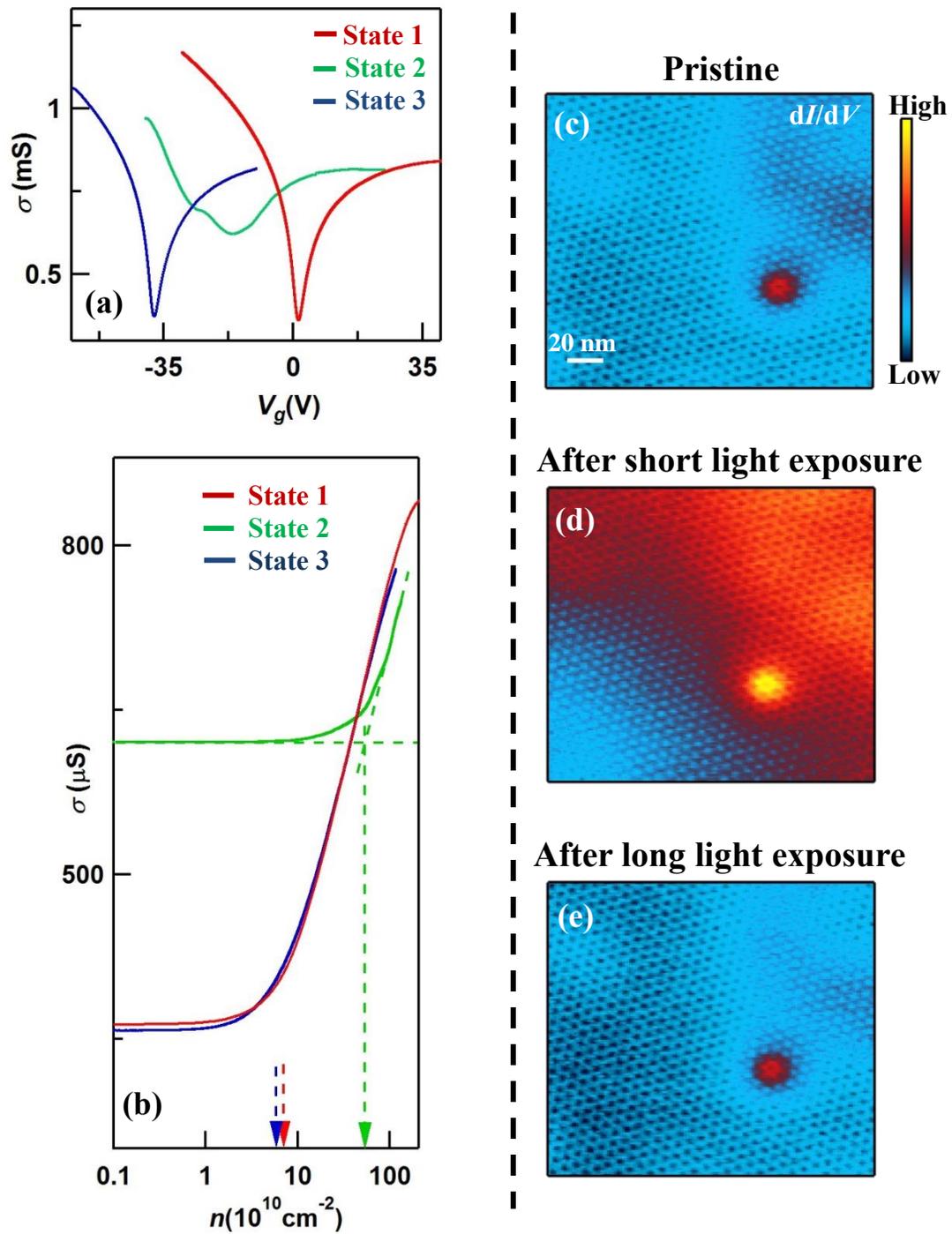

**Fig. 3**

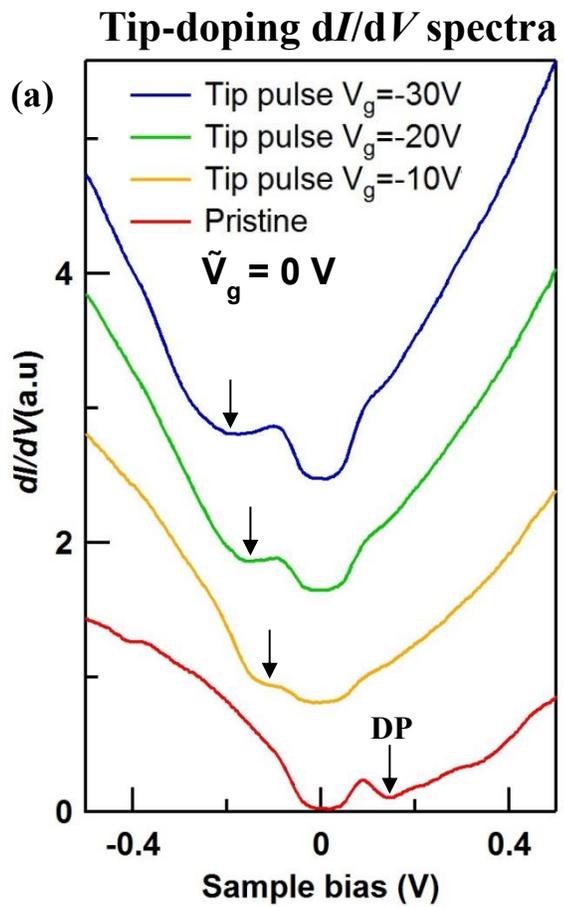

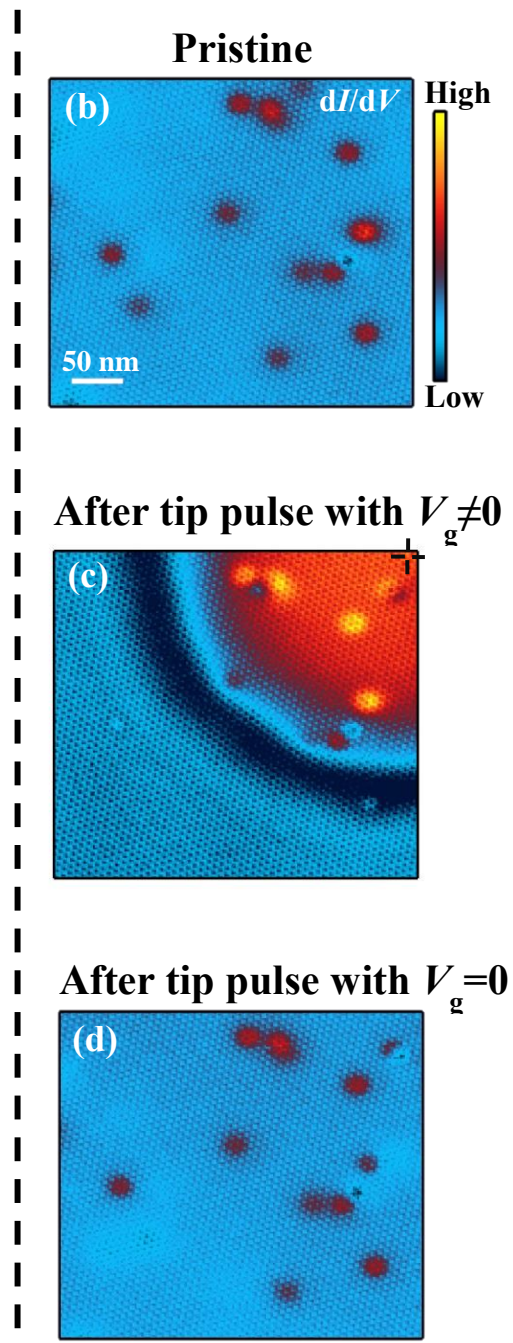

Fig. 4

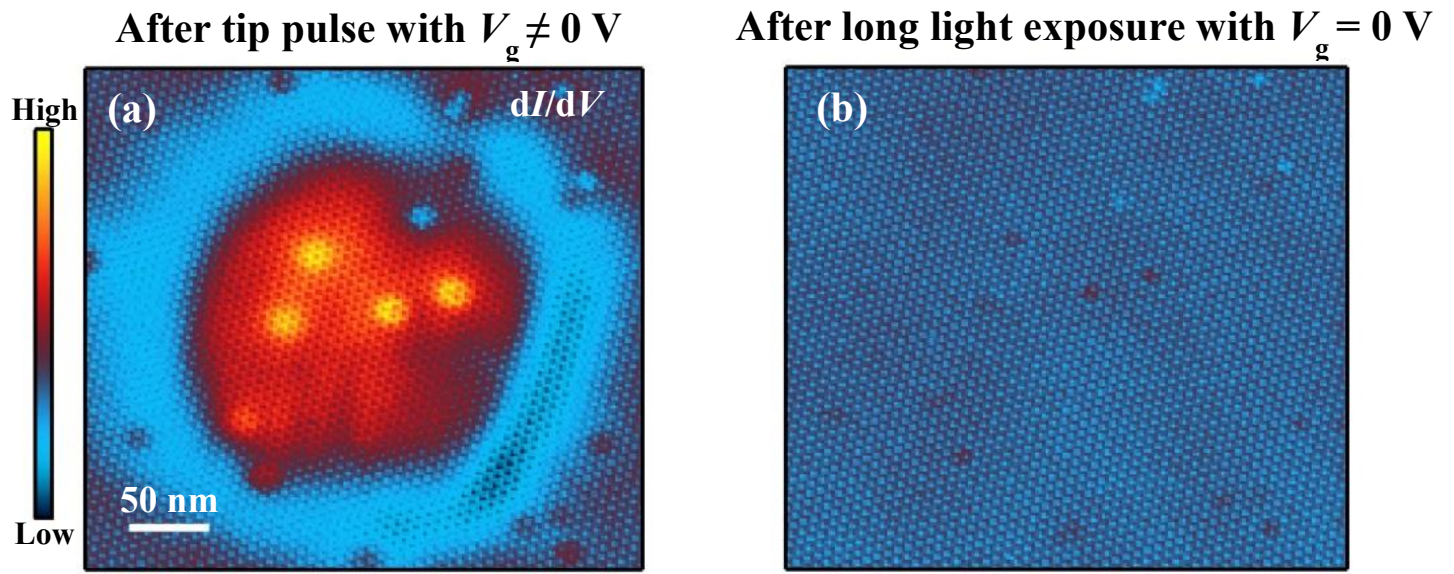